\documentclass[prd,aps,floats,twocolumn,showpacs]{revtex4}
\usepackage[dvips]{epsfig}

\begin{document}

\title{Normalized entropy density of the 3D 3-state Potts model}

\author{Alexei Bazavov$^{\rm \,a,b}$ and Bernd A. Berg$^{\rm \,a,b}$}

\affiliation{ $^{\rm \,a)}$ Department of Physics, Florida State
University, Tallahassee, FL 32306-4350, USA\\
$^{\rm \,b)}$ School of Computational Science,
Florida State University, Tallahassee, FL 32306-4120, USA}

\date{Feb 17, 2007}

\begin{abstract}
Using a multicanonical Metropolis algorithm we have performed Monte
Carlo simulations of the 3D 3-state Potts model on $L^3$ lattices
with $L=20$, 30, 40, 50. Covering a range of inverse temperatures
from $\beta_{\min}=0$ to $\beta_{\max}=0.33$ we calculated the
infinite volume limit of the entropy density $s(\beta)$ with its
normalization obtained from $s(0)=\ln 3$. At the transition temperature
the entropy and energy endpoints in the ordered and disordered phase
are estimated employing a novel reweighting procedure. We also evaluate
the transition temperature and the order-disorder interface tension.
The latter estimate increases when capillary waves are taken into
account.
\end{abstract}
\maketitle

\section{Introduction}

The 3D 3-state Potts model plays a role in our understanding of the
properties of the deconfining phase transition and the structure of
QCD. The high-temperature vacuum of QCD is characterized by ordered
Polyakov loops which behave similarly as spins in the low temperature
phase of the 3D 3-state Potts model. The Polyakov loop serves as an
order parameter of SU(3) pure gauge theory with its symmetry determined
by the $Z_3$ center of the gauge group, and the phase transition is
weakly first order. This maps naturally onto the order parameter
(magnetization) of the 3D 3-state Potts model \cite{SY82}. For
this and other reasons a number of numerical studies of the model
were performed \cite{GKB89,FMOU90,ABV91,Schm94,JV97,KaSt00}.

In this paper we use the multicanonical technique \cite{BN92}
to calculate the properly normalized entropy all the way from $\beta=0$
to across the phase transition. Besides this new result a number of
additional estimates (see the abstract), are obtained and compared
with the literature. Our paper is organized as follows: In section
\ref{sec_notation} we introduce basic notation and observables of
interest, in section \ref{sec_sim_res} we present our simulations
and conclusions are given in the final section \ref{sec_sum}.

\section{Notation and Preliminaries \label{sec_notation}}

We simulate the $q=3$ Potts model with the energy function
\begin{equation} \label{E_Potts}
  E\ =\  2 \sum_{\langle ij\rangle}
  \left(\frac{1}{q}-\delta_{q_iq_j}\right)
\end{equation}
where the sum is over the nearest neighbors of a 3D cubic lattice of
size $L^3$. The spins $q_i$ of the system take the values $q_i=0,\dots,
q-1$.  The factor of two and the term $1/q$ are introduced to match for
$q=2$ with Ising model conventions~\cite{BBook}.

Our simulations are carried out in a multicanonical ensemble
\cite{BN92}, covering an inverse temperature range from $\beta=
\beta_{\min}=0$ (infinite temperature) to $\beta=\beta_{\max}=0.33$
below the phase transition temperature. Instead of relying on a
recursion (see, e.g., \cite{BBook}), the multicanonical parameters
were extracted by finite size (FS) extrapolations from smaller to
larger system, which is efficient when the FS behavior is controllable.
Frequent excursions into the disordered phase all the way to $\beta=0$
secure equilibration of the configurations around the transition and
in the ordered phase.

By the fluctuation-dissipation theorem the specific heat is
\begin{equation}\label{sph}
  C_v=\frac{(\beta)^2}{L^3}
  \left(\langle E^2 \rangle-\langle E \rangle^2\right)\ .
\end{equation}
We put $\beta$ in parenthesis, because the number 2 is later used as
a superscript of $\beta$. The energy density per site is defined by
\begin{equation}\label{ee2}
  e=\frac{1}{L^3}\langle E \rangle\ ,
\end{equation}
so that the latent heat is
\begin{equation}\label{de_def}
  \Delta e=e^+-e^-\ ,
\end{equation}
where $e^+$ and $e^-$ are the energy endpoints of the high ($+$)
and low ($-$) temperature phase at the transition temperature
$T_c=1/\beta_c$. The entropy density is
\begin{equation}\label{entr_den}
  s=\beta\,(e-f)\ ,
\end{equation}
where $f$ is the free energy density. As the free energy is continuous
at the phase transition the entropy gap across the phase transition is
\begin{equation}\label{Sgap}
  \Delta s=\beta_c\,\Delta e\,.
\end{equation}
To determine the entropy and energy gaps from data on finite lattices
we follow Ref.~\cite{CLB86} and study the scaling of the specific heat
maxima $C_{\max}(L)$. The leading order coefficient $a_2$ of the fit
\begin{equation}\label{CmaxFS}
  C_{\max}(L)=a_1+a_2\,L^3
\end{equation}
is related to the latent heat by
\begin{equation} \label{Ea2}
  a_2=\frac{(\beta_c)^2\,(\Delta e)^2}{4}
\end{equation}
and using (\ref{Sgap}) gives
\begin{equation} \label{Sa2}
  \Delta s=2\sqrt{a_2}\ .
\end{equation}
The entropy densities in the disordered ($+$) and ordered ($-$) phase at
the transition temperature are defined by
\begin{eqnarray}\label{splus_def}
  s^+ &=& \beta_c\,(e^+-f(\beta_c))\,,\\
  s^- &=& \beta_c\,(e^--f(\beta_c))\ .
\end{eqnarray}
Calculation of the endpoints from Monte Carlo (MC) data faces some
technical difficulties, which are overcome in section~\ref{subsec_entr}.

In a MC simulation of the Gibbs canonical ensemble the entropy of the
system is only determined up to an additive constant, whereas in our
multicanonical simulation this constant is determined by the known
normalization at $\beta=0$:
\begin{equation}\label{S0}
  S_0=\ln\left( 3^{L^3} \right)~~{\rm and}~~
  s_0=\frac{S_0}{L^3}=\ln 3\simeq 1.098612\ .
\end{equation}

\section{Simulation results\label{sec_sim_res}}

In Table~\ref{tab_3DP3q} we list the lattice sizes used, number of
sweeps (sequential updates of the lattice for which each spin is
touched once) performed with the multicanonical Metropolis algorithm
and the number of cycles
$$ (\beta_e\le\beta_{\min}) \to (\beta_e\ge\beta_{\max}) \to
   (\beta_e\le\beta_{\min})\,, $$
the Markov process performed during the production run, where $\beta_e$
is the effective energy-dependent $\beta$ of the multicanonical
procedure. Data are energy histograms, which are collected over the
statistics given by the second number in the production statistics
column (sweeps) of Table~\ref{tab_3DP3q}. This is repeated for 32
bins, the first number in this column. The data analysis relies on
converting the bins into jackknife bins, which are then reweighted
to canonical ensembles using the logarithmic coding of~\cite{BBook}.
Error bars are obtained by performing analysis calculations for each
jackknife bin.

\begin{table}[ht]
\caption{Statistics of multicanonical simulations
of the 3-state Potts model on $L^3$
lattices.} \label{tab_3DP3q}
\medskip \centering
\begin{tabular}{|c|c|c|}   \hline
 $L$ & sweeps                    & cycles\\ \hline
\ 20~&$\ 32\times 1.2\cdot 10^5$ & \ 59  \\ \hline
\ 30 &$\ 32\times 5.2\cdot 10^5$ & \ 71  \\ \hline
\ 40 &$\ 32\times 1.5\cdot 10^6$ & \ 73  \\ \hline
\ 50 &$\ 32\times 6.0\cdot 10^6$ &  131  \\ \hline
\end{tabular} \end{table}

\subsection{Interface tension and specific heat maxima
            \label{subsec_interrface}}

\begin{figure}[t] \begin{center}
\epsfig{figure=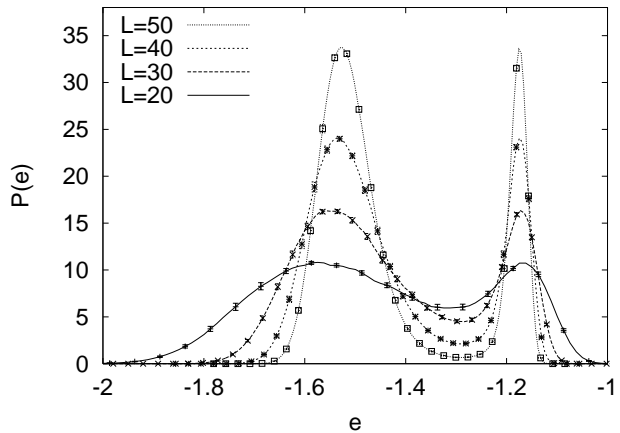,width=\columnwidth} 
\caption{Histograms of the energy density reweighted to equal heights
on various lattices.} \label{fig_eq_hists}
\end{center} \end{figure} 

\begin{table}[tf]
\caption{First inverse pseudo-transition temperature and the
interface tension on $L^3$ lattices.} \label{tab_3DP3q_beta}
\medskip \centering
\begin{tabular}{|c|c|c|c|}   \hline
$L$ & $\beta^1_{pt}$ & 2$\sigma_{od}$ & $C_{\max}$ \\ \hline
 20 & 0.275311 (33)  & 0.00152 (14)   & 26.27 (51) \\ \hline
 30 & 0.275284 (11)  & 0.001440 (49)  & 62.49 (88) \\ \hline
 40 & 0.2752853 (59) & 0.001522 (37)  & 136.4 (1.5)\\ \hline
 50 & 0.2752838 (23) & 0.001574 (22)  & 263.2 (1.7)\\ \hline
\end{tabular} \end{table}

Strictly speaking there are no phase transitions on finite lattices.
But one can define pseudo-transition temperatures $T_{pt}(L)=1/
\beta_{pt}(L)$, which agree up to $1/L^3$ corrections with $T_c$.
We give a first definition, $\beta^1_{pt}(L)$, as the $\beta$
values which reweight the double-peaked histogram at the transition
to {\it equal heights}. These histograms are presented in
Fig.~\ref{fig_eq_hists} and corresponding $\beta^1_{pt}(L)$
values are given in Table~\ref{tab_3DP3q_beta}. Another definition,
$\beta^2_{pt}(L)$, will be introduced later. For the first-order
phase transition the FS behavior of these pseudo-transition
definitions is (up to higher orders in $1/L^3$)
\begin{equation}\label{betaFS}
  \beta^i_{pt}(L)=\beta^i_c+\frac{b^i}{L^3}\,,~~i=1,2\,.
\end{equation}
Fitting the $\beta^1_{pt}(L)$ values of Table~\ref{tab_3DP3q_beta} to
this equation yields
\begin{equation}\label{beta1c}
  \beta^1_c=0.2752827\,(29),\,\,\,\,\,b^1=0.15\,(22)
\end{equation}
with a goodness of fit \cite{BBook} $Q=0.89$. Within statistical
errors the coefficient $b^1$ is zero, indicating that the FS
corrections to $\beta^1_c$ are mild.

For $L\to\infty$ the interface tension between ordered and disordered
phases is~\cite{Bi82}
\begin{equation}\label{sigma}
  2 \sigma_{od}(L) = \frac{1}{L^2}\ln
  \left(\frac{P_{\max}(L)}{P_{\min}(L)}\right)
\end{equation}
where $P_{\max}(L)$ represents the value of the maxima when the energy
histogram is reweighted to equal heights and $P_{\min}(L)$ the minimum
in between the peaks (Fig.~\ref{fig_eq_hists}). The results are
contained in Table~\ref{tab_3DP3q_beta}. A fit to the form
\begin{equation}\label{sigma_fit0}
  2\,\sigma_{od}(L) = 2\sigma_{od} + \frac{c}{L^2}
\end{equation}
gives
\begin{equation}\label{sigma_od0}
  2\sigma_{od}=0.001602\,(35)\,,\qquad c=-0.100\,(52)
\end{equation}
with $Q=0.22$. This is consistent with previous literature:
$2\sigma_{od}=0.001568\,(52)$ of \cite{Schm94} ($Q=0.59$) and
$2\sigma_{od}=0.00163\,(2)$ of \cite{JV97} ($Q=0.49$), where the
$Q$ values are now from Gaussian difference tests \cite{BBook} with
our fit. However, if one includes capillary waves \cite{BZ,GF,Mo}
the fit form becomes, see Eq.~(16) of \cite{BNB},
\begin{equation}\label{sigma_fit}
  2\,\sigma_{od}(L) + \frac{\ln(L)}{2L^2} =
  2\sigma_{od} + \frac{c}{L^2}
\end{equation}
and we obtain a different consistent estimate:
\begin{equation}\label{sigma_od}
  2\sigma_{od}=0.001806\,(35)\,,\qquad c=1.379\,(52)
\end{equation}
with $Q=0.96$. \medskip

Table~\ref{tab_3DP3q_beta} collects also our estimates of the maxima of
the specific heat $C_{\max}(L)$. Fitting them all to Eq.~(\ref{CmaxFS})
gives the rather low goodness of fit $Q=0.028$. Following standard
practice we omit therefore the smallest ($L=20$) lattice from the
fit and find
\begin{equation}\label{a1a2}
  a_1=7.0\,(1.3)~~~{\rm and}~~~a_2=0.002044\,(20)
\end{equation}
with $Q=0.28$.

\subsection{Critical temperature, entropy and energy\label{subsec_entr}}

\begin{figure}[-t] \begin{center}
\epsfig{figure=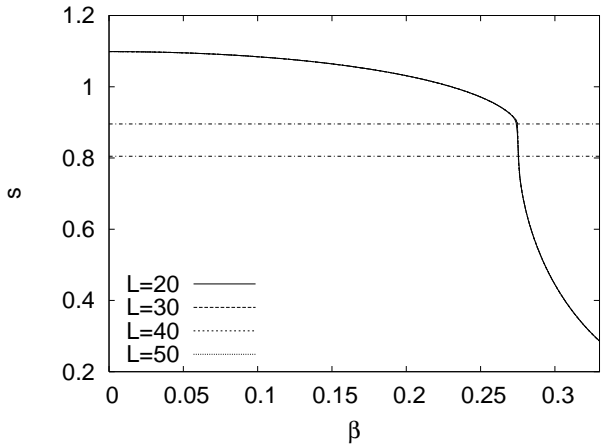,width=\columnwidth} 
\caption{Entropy density of the 3D 3-state Potts model on different
lattices. Horizontal lines represent $s^+$ and $s^-$ values. FS
effects are not visible on the scale of this figure.}
\label{fig_entropy0}
\end{center} \end{figure} 

\begin{figure}[-t] \begin{center}
\epsfig{figure=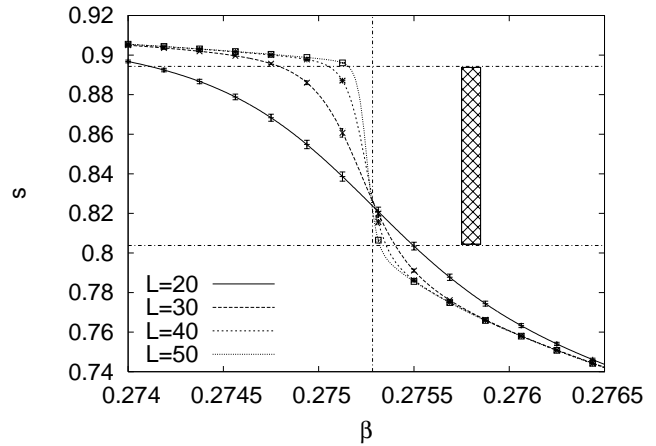,width=\columnwidth} 
\caption{Entropy density in the vicinity of the phase transition. The
height of the bar on the right represents the entropy gap $\Delta s$.
The horizontal lines are $s^+$ and $s^-$ and the vertical line is
$\beta^2_c$.}\label{fig_entropygap}
\end{center} \end{figure}

The entropy density from our multicanonical simulations is shown in
Fig.~\ref{fig_entropy0}. Error bars and FS corrections are not visible
on the scale of this figure and the vicinity of the phase transition
is enlarged in Fig.~\ref{fig_entropygap}.

To determine the endpoints of the entropy and energy on the disordered
and ordered side of the transition, we introduce a second definition
of inverse pseudo-transition temperatures, $\beta^2_{pt}(L)$, which
differ from the equal heights definition in crucial details. We adjust
the weights of the double peak histograms so that
\begin{equation}\label{beta0p}
   e_L(\beta^2_{pt}) = \frac{1}{2} \left[ e_L^+(\beta^2_{pt})
                     + e_L^-(\beta^2_{pt}) \right]
\end{equation}
holds, where the {\it central energy density} $e_L(\beta^2_{pt})$
is the expectation value at $\beta^2_{pt}(L)$ and
$e_L^{\pm}(\beta^2_{pt})$ are the locations of the maxima of the
double peak histogram at $\beta^2_{pt}(L)$, $e_L^+$ on the disordered
and $e_L^-$ on the ordered side of the transition. The idea behind
this construction is to ensure that the energy endpoints (\ref{de_def})
are positioned symmetrically about the central energy density:
\begin{equation}\label{epmsym}
  e_L^{\pm} = e_L(\beta^2_c) \pm \frac{1}{2}\,\Delta e_L\,.
\end{equation}
In the following we call this procedure {\it equal distances}
reweighting. Using (\ref{entr_den}) and (\ref{Sgap}) one finds
that
\begin{equation}\label{spmsym}
  s_L^{\pm} = s_L(\beta^2_c) \pm \frac{1}{2}\,\Delta s_L
\end{equation}
holds as well.

\begin{figure}[-t] \begin{center}
\epsfig{figure=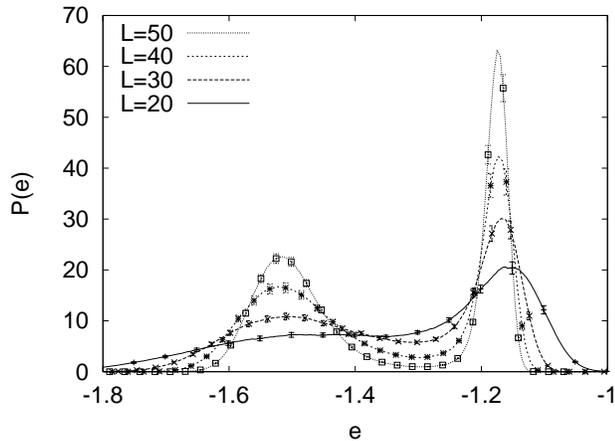,width=\columnwidth} 
\caption{Histograms of the energy density reweighted to equal distances
on various lattices.} \label{fig_eq_dists}
\end{center} \end{figure} 

\begin{table}[tf]
\caption{Second inverse pseudo-transition temperature, central energy
and and entropy densities on $L^3$ lattices.} \label{tab_3DP3q_beta2}
\medskip \centering
\begin{tabular}{|c|c|c|c|c|}   \hline
$L$& $\beta^2_{pt}$&$e_L(\beta^2_{pt})$&$s_L(\beta^2_{pt})$
                                       &$f_L(\beta^2_{pt})$\\ \hline
30 &0.275166\ (13)&$-1.336~$ (14)&0.8522 (37)&$-4.43275~$ (14)\\ \hline
40 &0.2752369 (52)&$-1.345~$ (11)&0.8497 (29)&$-4.431840$ (62)\\ \hline
50 &0.2752508 (22)&$-1.3444$ (51)&0.8498 (14)&$-4.431652$ (26)\\ \hline
\end{tabular} \end{table}

The histograms reweighted to equal distances are shown in
Fig.~\ref{fig_eq_dists} and the corresponding $\beta^2_{pt}(L)$ are
collected in Table~\ref{tab_3DP3q_beta2}. For the $L=20$ histogram of
the figure $\beta=0.275311$ is used just to illustrate that the double
peak disappears before condition (\ref{beta0p}) gets fulfilled. This
is the reason why there are no $L=20$ values in the table. Fitting
the $\beta^2_c(L)$ values of Table~\ref{tab_3DP3q_beta2} to
(\ref{betaFS}) yields
\begin{equation}\label{beta2c}
  \beta^2_c=0.2752724\,(44),\,\,\,\,\,b^2=-2.61\,(40)
\end{equation}
with a goodness of fit $Q=0.21$. The estimates (\ref{beta1c}) and
(\ref{beta2c}) are at the edge of being consistent ($Q=0.051$ for
the Gaussian difference test). Our second estimate for $\beta_c$
is less accurate than our first, apparently because the location
of the newly defined maxima of the energy probability density is
less stable than for the equal heights definition. Also FS
corrections to the second estimate are not negligible as those to
the first. In absolute numbers the difference is less than 1 in
the fifth significant digits, so it does not propagate into the
$\beta_c$ factor used for the far less accurate latent heat. Still
the second definition is needed to enable estimates of $e^{\pm}$
and $s^{\pm}$.

To give a combined estimate of $\beta_c$, we average the $\beta^1_c$
and $\beta^2_c$ estimates weighted by their inverse variances and
keep the error bar \cite{procedure} of the more accurate $\beta^1_c$
result to obtain
\begin{equation}\label{betac}
  \beta_c=0.2752796\,(29)\,.
\end{equation}
For comparison, the weighted average of \cite{JV97} on lattices
$L=10,\,\dots\, ,36$ is $\beta_c=0.2752825(50)$ and the Gaussian
difference test with our value gives $Q=0.62$.

To calculate $\Delta s$ and $\Delta e$ we define jackknife estimators
on finite lattices for them:
\begin{equation} \label{dsj}
  \Delta s_L = 2\,\sqrt{\frac{C_{\max}(L)}{L^3}}\,.
\end{equation}
and
\begin{equation} \label{dej}
  \Delta e_L = \frac{2}{\beta^2_{pt}(L)}\,
            \sqrt{\frac{C_{\max}(L)}{L^3}}\ .
\end{equation}
Their values will be needed for combination with jackknifed central
entropy and energy values, so error bars for $e^{\pm}$ and $s^{\pm}$
can be obtained. Only $L\ge 30$ lattices are used as we
have already seen for $a_2$ that the smallest lattice spoils the fit.
Estimates of the means and their error bars follow in the usual jackknife
way. Fitting them with $1/L^3$ corrections, the entropy gap is
\begin{equation} \label{sgap}
  \Delta s=0.09045\,(41)\ .
\end{equation}
It is represented by the height of the filled bar on the right of
Fig.~\ref{fig_entropygap}. In the same way we find for the latent heat
\begin{equation}\label{de_cmax}
  \Delta e=\frac{\Delta s}{\beta_c}=0.3286\,(15)\ .
\end{equation}
This value is somewhat larger than $\Delta e=0.32320\,(94)$ of
\cite{JV97} ($Q=0.002$) and in agreement with $\Delta e=0.3291\,(33)$
of \cite{ABV91} where additional data of \cite{FMOU90} on lattices up
to $L=48$ was taken into account ($Q=0.89$). However, these results
depend still beyond the range of one error bar on the order in which the
functions are evaluated. Using (\ref{Sa2}) and (\ref{Ea2}) to estimate
the gaps from $a_2$ of Eq.~(\ref{a1a2}), one finds $\Delta s = 0.08933
\,(38)$ and $\Delta e = 0.3245\,(14)$. This $\Delta e$ value is
consistent with both results of the previous literature as it is
in the middle of them.

The central energy (\ref{beta0p}), entropy and free energy density
values (all defined at $\beta^2_{pt}$) are listed in
Table~\ref{tab_3DP3q_beta2}. From
\begin{equation}
  e_L(\beta^2_{pt}) = e(\beta^2_c)+\frac{c}{L^3},
\end{equation}
we find the infinite volume extrapolation
\begin{equation}
  e(\beta^2_c) = -1.3470\,(74)\,,\qquad Q=0.82\,.
\end{equation}
Similarly we determine
\begin{equation}\label{s0fit}
  s(\beta^2_c) = 0.8491\,(21)\,,\qquad Q=0.82\,,
\end{equation}
and the corresponding (\ref{entr_den}) free energy density
\begin{equation}\label{f0fit}
  f(\beta^2_c) = -4.431364\,(50)\,,\qquad Q=0.18\,.
\end{equation}
The statistical errors of $f_L(\beta^2_{pt})$ are quite small due
to correlations between $e_L(\beta^2_{pt})$ and $s_L(\beta^2_{pt})$.

\begin{table}[tf]
\caption{Results for $e^{\pm}_L$ (\ref{epmsym}) and $s^{\pm}_L$
(\ref{spmsym}).} \label{tab_3DP3q_pm}
\medskip \centering
\begin{tabular}{|c|c|c|c|c|}   \hline
$L$& $e^+_L$       & $e^-_L$       & $ s^+_L$   & $s^-_L$    \\ \hline
30 & $-1.161~$ (13)& $-1.511~$ (15)& 0.9003 (36)& 0.8041 (39)\\ \hline
40 & $-1.177~$ (10)& $-1.513~$ (12)& 0.8958 (28)& 0.8035 (31)\\ \hline
50 & $-1.1777$ (50)& $-1.5111$ (53)& 0.8957 (14)& 0.8039 (15)\\ \hline
\end{tabular} \end{table}

From (\ref{epmsym}) and (\ref{spmsym}) we find finite lattice
estimators $e_L^{\pm}$ and $s_L^{\pm}$, which are collected in
Table~\ref{tab_3DP3q_pm}. Fits with $1/L^3$ corrections give:
\begin{eqnarray} \label{ep}
  e^+ &=& -1.1826\,(73)\,,\qquad Q=0.74\,, \\ \label{em}
  e^- &=& -1.5112\,(79)\,,\qquad Q=0.88\,, \\ \label{sp}
  s^+ &=&  0.8943\,(21)\,,~~\qquad Q=0.73\,, \\ \label{sm}
  s^- &=&  0.8038\,(22)\,,~~\qquad Q=0.90\,.
\end{eqnarray}
The values of $s^+$ and $s^-$ are shown in Figs.~\ref{fig_entropy0}
and~\ref{fig_entropygap} as horizontal lines. The vertical line in
Fig.~\ref{fig_entropygap} represents the inverse transition temperature
(\ref{beta2c}).

\section{Summary and Conclusions \label{sec_sum}}

The main result of this work is the entropy density with proper
normalization, shown in Fig.~\ref{fig_entropy0}, where $s_0=\ln 3$.
The transition region is enlarged in Fig.~\ref{fig_entropygap}.
The values of the entropy density $s^+$ (\ref{sp}) on the disordered
and $s^-$ (\ref{sm}) on the ordered side of the transition in \% of
$s_0$ are, respectively, 82\% and 73\%. Thus, there are 3 states per
site at infinite temperature and effectively 2.45 states on the high-
and 2.24 states on the low-temperature side of the phase transition.
Our corresponding $e^{\pm}$ estimates are given in Eqs.~(\ref{ep})
and (\ref{em}).

Other results are $\beta_c$ (\ref{betac}), the free energy density
at the critical point (\ref{f0fit}), the entropy gap (\ref{sgap}),
the latent heat (\ref{de_cmax}) and the interface tension
(\ref{sigma_od}). It is notable that the inclusion of capillary
waves enhances the estimate of the interface tension by more than
10\%.

\acknowledgments
This work was in part supported by the DOE grant DE-FG02-97ER41022.


\begin{thebibliography}{19}
\bibitem{procedure} This error bar should not be reduced when combining
$\beta^1_c$ and $\beta^2_c$, because they rely on the same simulations.
\end{thebibliography}
\end{document}